\documentclass{iau}

\usepackage{graphicx}

\title[IAUS291.~~Pulsar Wind Nebulae: 
 What do they tell us about pulsars?] 
{Pulsar Wind Nebulae: \\On their growing diversity and association with highly magnetized neutron stars}

\author[S. Safi-Harb]
{Samar Safi-Harb\thanks{Canada Research Chair}}

\affiliation{Department of Physics \& Astronomy, University of Manitoba, Winnipeg, MB,  Canada \\ email: {\tt samar@physics.umanitoba.ca}\\[\affilskip]}

\pubyear{2012}
\volume{291}  
\jname{\mbox{Neutron Stars and Pulsars: Challenges and Opportunities after 80 years}}
\editors{J. van Leeuwen, ed.} 
\begin{document}

\maketitle

\begin{abstract}
The 1968 discovery of the Crab and Vela pulsars in their respective supernova remnants (SNRs) confirmed Baade and Zwicky's 1934 prediction that supernovae form neutron stars. Observations of Pulsar Wind Nebulae (PWNe), particularly with the Chandra X-ray Observatory, have in the past decade opened a new window to focus on the neutron stars' relativistic winds, study their interaction with their hosting SNRs, and find previously missed pulsars. While the Crab has been thought for decades to represent the prototype of PWNe, we now know of different classes of neutron stars and PWNe whose properties differ from the Crab. In this talk, I review the current status of neutron stars/PWNe-SNRs associations, and 
highlight the growing diversity of PWNe with an X-ray eye on their association with highly magnetized neutron stars.
I conclude with an outlook to future high-energy studies.
\end{abstract}

\firstsection 
\section{Brief history and  current status of pulsar--PWN--SNR associations}
Shortly after the discovery of the neutron particle by Chadwick (1932), Baade and Zwicky (1934) made the seminal prediction that neutron stars are born in supernovae.
Very little was known about the manifestation of these objects until Jocelyn Bell and Antony Hewish 
discovered the first pulsating star with a period of 1.3 sec (Hewish et al. 1968), now known as PSR B1919+21.
In the same year, the Crab (Staelin \& Reifenstein  1968, Lovelace et al. 1968)
and Vela (Large et al. 1968) pulsars were discovered in their respective supernova remnants (SNRs)
confirming the 1934 prediction for the association of neutron stars with supernovae.

On the theoretical side, Pacini (1967)\footnote[2]{a pioneer in PSR theory who recently passed away
(25 Jan. 2012);  a few months before PSR discovery, he predicted that 
neutron stars could release their rotational energy through jets.},
 Gold (1968),  Pacini \& Salvati (1973), and Rees \& Gunn (1974)
introduced the theory that the emission from a neutron star is powered by its rotational energy loss, $\dot{E}$=4$\pi^2 I \dot{P}/P^3$ (with $P$ and $\dot{P}$ being the rotation period and its derivative),
with the pulsar generating a magnetized particle wind whose ultra-relativistic electrons and
positrons emit synchrotron radiation across the electromagnetic spectrum. As this wind encounters its confining surrounding medium, it gets shocked
forming a nebula referred to as a pulsar wind nebula (PWN, also known as $plerion$).

PWNe, being the bubbles inflated by the spin-down energy of the pulsar, 
 have proven to be excellent pathfinders for pulsar discovery. Even in the absence of direct pulsar detection, their emission immediately implies the presence of a neutron star
whose properties ($\dot{E}$, $P$, $\dot{P}$, characteristic age $\tau_c$=$P$/2$\dot{P}$,  and dipole magnetic field $B$=3.2$\times$10$^{19}$~$(P \dot{P})^{1/2}$~G)
can be directly inferred from the properties of the PWN (see Gaensler \& Slane 2006 and Kargaltsev \& Pavlov 2008 for reviews).

Up to the early 1990s only a handful of neutron stars were known to be associated with SNRs. The number of associations has however significantly increased with the synergy of radio and X-ray studies,
 particularly with the imaging and spectroscopic capabilities of the $ROSAT$ and $ASCA$ missions in the 1990s, and since 2000 with the $Chandra$ and \textit{XMM-Newton}
missions. $ASCA$'s coverage in the hard X-ray band ($E$$\gtrsim$2~keV) has been particularly instrumental in identifying PWNe inside SNRs,
therefore revealing new pulsars following targeted radio and X-ray pulsation searches.
$Chandra$'s superb imaging resolution has opened a new window to image these fascinating objects through revealing high-resolution, arcsecond-scale, structures associated with the deposition of the pulsar's wind energy
into its surroundings, and to pin down their powering engines.

Currently out of the  309~known Galactic SNRs, 103 are associated with a neutron star or a neutron star candidate, with 85 being identified as a pulsar.
A PWN is detected or suggested in 87 cases, with 62 SNRs  associated with both a PWN and a neutron star or pulsar (see Ferrand \& Safi-Harb 2012 and these proceedings).

\section{The growing diversity of neutron stars and PWNe}
One of the major discoveries advanced in the past decade, thanks to the synergy of radio and X-ray observations,
is the growing diversity of neutron stars. These include, in addition to the rotation-powered pulsars (RPP) like the Crab, 
the Anomalous X-ray pulsars (AXPs) and Soft Gamma-ray Repeaters (SGRs) dubbed as `magnetars' with surface dipole magnetic fields, $B$, exceeding the
quantum electrodynamic threshold $B_{QED}$=4.4$\times$10$^{13}$~G, the high-magnetic field radio pulsars (HBPs) with magnetic fields intermediate between the Crab-like pulsars and magnetars,
the Rotating Radio Transients (RRATs), and the Central Compact Objects (CCOs) inside SNRs currently believed to be `anti-magnetars' (see Mereghetti 2008,
Ng \& Kaspi 2011, Keane et al. 2011, and Gotthelf \& Halpern 2008, respectively, for reviews on these neutron star classes).
Among the 103 neutron star--SNR associations, 10 SNRs are associated with magnetars or magnetar candidates (including 6 $proposed$ associations), 2 SNRs are securely associated with HBPs,
13 SNRs are associated or proposed to be associated with CCOs/CCO candidates,
with the remaining SNRs associated with RPPs or candidate neutron stars.

An outstanding question in this field is whether these apparently different classes of neutron stars are linked. It is  now clear (e.g. from the discovery of radio emission from a few magnetars,
magnetar-like activity of the HBP J1846--0258 in the SNR~Kes~75, and the existence of low-B magnetars like SGR~0418+5729) that the dipole magnetic field 
is not the sole factor determining their observational properties. One big unknown is their environment and progenitors (see Safi-Harb \& Kumar, these proceedings). 
Studies and/or searches for any associated PWNe can help shed light on these questions and provide further clues on the nature of their emission mechanism and their confining environment.

In fact, SNR studies, conducted in parallel with neutron stars studies, have also revealed new classes of PWNe with properties unlike the Crab nebula long thought to be the prototype PWN.
These include PWNe expanding into bubbles blown by their progenitors, evolved objects powering X-ray nebulae that are offset from their radio counterparts, 
or compact X-ray PWNe powered by pulsars with likely highly magnetized winds. 
Coordinated targeted radio and X-ray studies of such unusual PWNe are revealing 
pulsar candidates awaiting discovery, or interesting pulsars  such as the highly energetic 24~ms pulsar recently discovered in the SNR G76.9+1.0 (Arzoumanian et al. 2011).
The diversity arising from such `unusual'  and evolved PWNe has been highlighted elsewhere (Safi-Harb 2012).
In this paper, I review the growing evidence for PWNe around the highly magnetized neutron stars and the clues they offer about their powering engines.

\section{Magnetar Wind Nebulae?}

\begin{figure}[b]
\center
\includegraphics[trim=0 10 0 0, clip, width=0.6\textwidth]{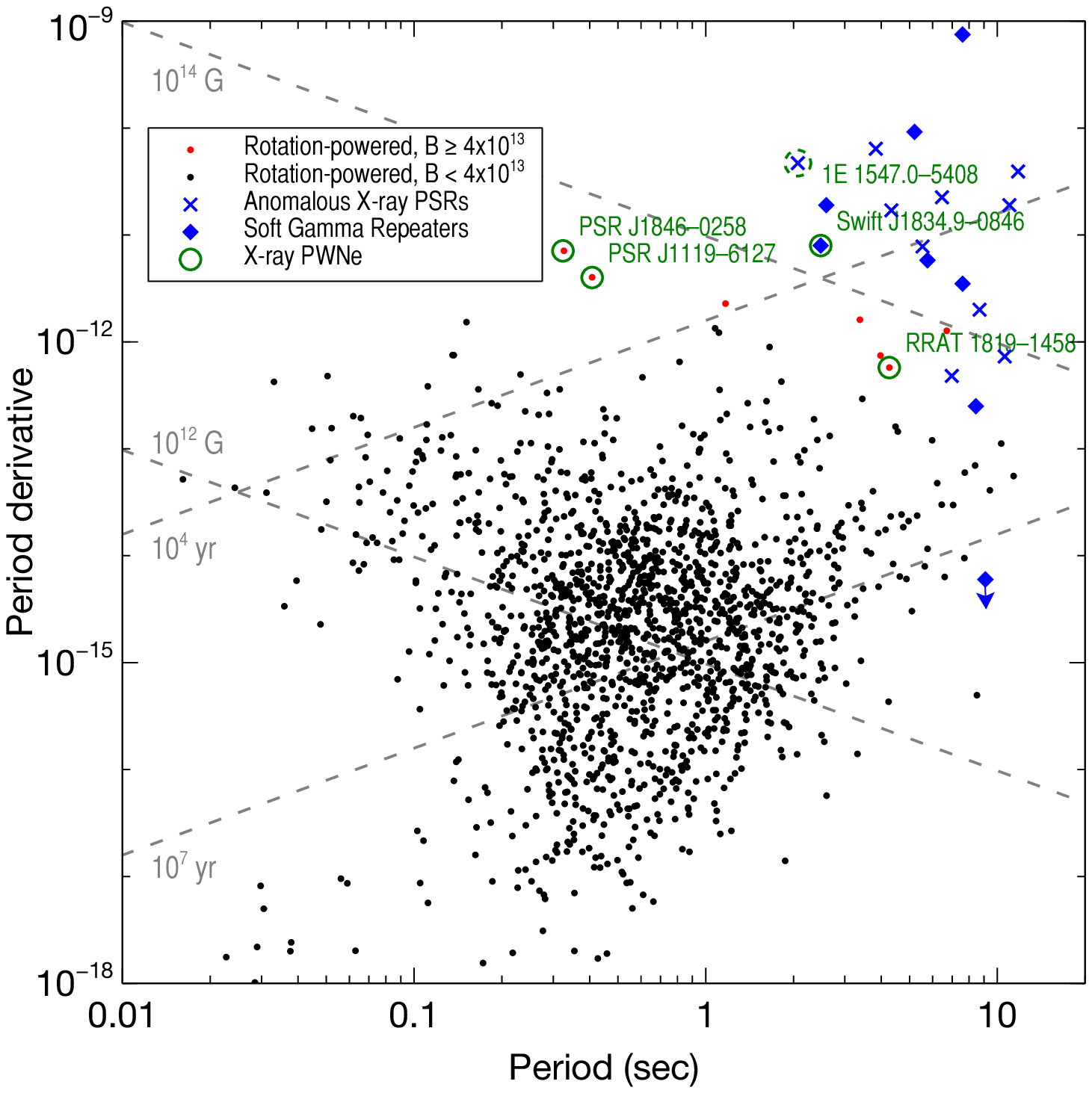}
\caption{$P$--$\dot{P}$ diagram for the rotation-powered pulsars (RPPs) and highly magnetized neutron stars (HBPs, AXPs and SGRs),
 highlighting the X-ray PWNe detected (green circles) or claimed (dashed green circle) around highly magnetized neutron stars with $B$$\gtrsim$4$\times$10$^{13}$~G. }
\end{figure}

Whether highly magnetized neutron stars (magnetars and HBPs with $B$$\gtrsim$4$\times$10$^{13}$~G) should power PWNe
is an open and interesting question.
For PWNe to form, one generally needs a relativistic particle outflow in a strongly magnetized and confining medium.
Magnetars have been suggested to produce steady or post-outburst particle outflows (e.g. Harding et al. 1999). Evidence for 
an intermittent outflow came with the discovery of a radio nebula around SGR~1806--20 following its 2004 December 27 giant flare (Gaensler et al. 2005). 
It's not clear however whether steady magnetar outflows can power detectable PWNe.
Their presence is of particular interest since they reflect the energy
budget released over the pulsar's lifetime. In addition, their X-ray spectral properties
shed light on whether they are rotation- or magnetically-powered, thus providing further clues to their possible link to the classical RPPs.

One of the challenges in identifying PWNe around magnetars lies in the magnetars' relatively high X-ray luminosity combined with heavy interstellar absorption causing the formation of a dust scattering
halo around them. Furthermore, typically the X-ray luminosity of the PWNe associated with the RPPs is only a small fraction of their
spin-down energy ($L_x$$\sim$10$^{-5}$--10$^{-2}$$\dot{E}$).
For magnetars, the spin-down energy is very small, ranging from
$\sim$3$\times$10$^{29}$~erg~s$^{-1}$ for SGR~0418+5729 to $\sim$2$\times$10$^{35}$~erg~s$^{-1}$ for 1E~1547.0--5408, with typical values of $\sim$10$^{33}$~erg~s$^{-1}$.
Therefore, assuming a similarly small X-ray to spin-down luminosity ratio for steady PWNe around magnetars, very deep high-resolution X-ray observations will be needed to identify any associated PWN.
Furthermore, the highest $\dot{E}$ magnetars would be the more promising candidates for PWN searches.

\begin{figure}[tbh]
\center
\includegraphics[trim=0 30 0 15, clip, width=0.95\textwidth]{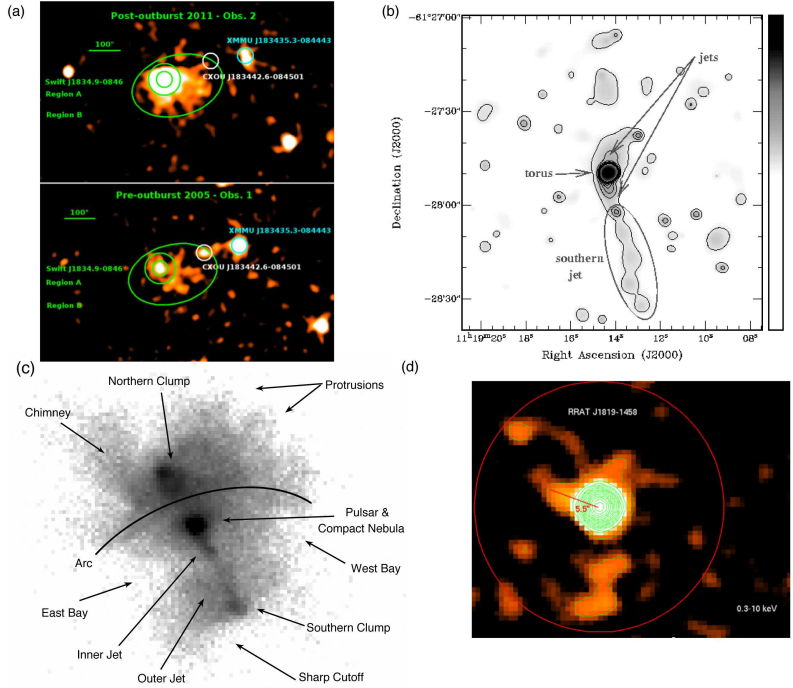}
\caption{X-ray PWNe around highly magnetized neutron stars: (a) the SGR Swift~J1834.9--0846 with \textit{XMM-Newton} showing the extended emission post- (top) and pre- (bottom) outburst observed in 2011, (b) the HBP J1119--6127 in SNR~G292.2--0.5 with $Chandra$ showing a compact PWN and a jet, 
(c) the youngest HBP J1846--0258 in SNR~Kes~75 with $Chandra$ revealing high-resolution structures and variability associated with the magnetar-like outburst observed in 2006,
and (d) the highest-$B$ RRAT~J1819--1458 showing a compact PWN with $Chandra$ (see also Camero et al., these proceedings). Figure adapted from figures published in Younes et al. 2012, Safi-Harb \& Kumar 2008, Ng et al. 2008,
and Rea et al. 2009. }
\end{figure}

So far, X-ray PWNe have been detected around the high-magnetic field pulsars J1119--6127, J1846--0258, and RRAT~J1819--1458 (the highest $B$ source among the RRATs) using $Chandra$, and
 the SGR Swift J1834.9--0846 using \textit{XMM-Newton}; see Table~1 for a summary of their properties, together with Figures~1 and 2 for their locations on the $P$-$\dot{P}$ diagram and their PWN morphology, respectively.
 Furthermore, a PWN has been suggested around AXP 1E~1547.0--5408 (Vink \& Bamba 2009), the largest
 spin-down luminosity magnetar; however a more recent study
showed that its emission is variable, correlated with the AXP's flux, and dominated by dust scattering (Olausen et al. 2011).

As shown in Table~1, these nebulae are compact and display a wide range of spectral indices and X-ray luminosities (or ratios with respect to their pulsar's spin-down luminosity). 
The compact PWN surrounding PSR~J1119--6127 is characterized by a hard power-law photon index (noting the jet's photon index is also hard, $\Gamma$=1.4$^{+0.8}_{-0.7}$) and a low X-ray to spin-down luminosity ratio, very similar to
the properties of PWNe associated with RPPs (Safi-Harb \& Kumar 2008). To date, this pulsar hasn't shown evidence for a magnetar-like burst.
PSR~J1846--0258,  an HBP with spin properties similar to PSR~J1119--6127 (although younger and not yet detected as a radio pulsar), has shown evidence for magnetar-like behaviour and variability in the PWN properties (Gavriil et al. 2008, Kumar \& Safi-Harb 2008, Ng et al. 2008). 
This so far remains the only HBP that has shown `schizophrenic' behaviour (i.e. RPP
and magnetar-like) suggesting that HBPs can be powered by both rotational energy and magnetic field decay (Camilo 2008). 

PSR~J1119--6127 and PSR~J1846--0258 are the only HBPs known to date to be securely associated with SNRs (as expected from their youth). They are  both characterized by a high $\dot{E}$
  and have a measured braking index yielding  a more accurate estimate for their actual ages. 
For PSR~J1119--6127,  the recently refined braking index measurement using more than 12 years of radio timing data  of $n$=2.684$\pm$0.002 (Weltevrede et al. 2011) implies
an upper limit on its age of 1.9~kyr, smaller than the estimated age of 4.2--7.1~kyr for its associated SNR, G292.2--0.5 (Kumar et al. 2012). This apparent age discrepancy can be 
attributed to a variable braking index for the pulsar, which has
recently also
shown some unusual timing characteristics at radio wavelengths.

The photon index for the PWNe associated with both above-mentioned HBPs is relatively hard and consistent with that observed in the RPPs' PWNe. 
For the RRAT and the Swift sources however,  their PWNe have a much steeper photon index, more similar to that observed in the (soft) power-law component of AXPs in the 
0.5--10 keV band, suggesting a similar population of particles emitting in X-rays.
The X-ray to spin-down luminosity of the three secure PWNe (excluding 1E 1547.0--5408 and with the exception of PSR~J1119--6127)  is relatively large in comparison to the RPPs. That, together with their relatively steep photon index and evidence of variability, $suggests$ that their X-ray luminosity is not entirely powered by rotation,  but possibly by some additional source of magnetic energy.
Energetically however, they can still be powered by rotation given that their X-ray luminosity is still smaller than their spin-down luminosity.

\begin{table}[b]
\caption{X-ray PWNe around high-B pulsars listed in the order of increasing $B$.}
\begin{tabular}{llccclccll}
\hline
PSR & $P$ (s) &  $B$  & $\dot{E}$   & $\tau_c$ & PWN & $\Gamma$ & $\frac{L_x}{\dot{E}}$  &  SNR? & Ref\\
 &  & \hspace*{-1mm}(10$^{13}$G) & (erg~s$^{-1}$) & (kyr) & extent &  & & & \\ \hline
 J1119--6127 & 0.408 &  4.1 & 2.3$\times$10$^{36}$& 1.7 & 6$\times$15$^{\prime\prime}$ & 1.1$^{+0.9}_{-0.7}$ & 5$\times$10$^{-4}$  & G292.2--0.5 & [1]  \\
J1846--0258  & 0.324 & 5 & 8.3$\times$10$^{36}$ & 0.7 &  40$^{\prime\prime}$ & 1.8$\pm$0.1 & 0.2--0.3 &  Kes~75  & [2] \\
J1819--1458 & 4.26 &  5 & 3$\times$10$^{32}$ & 117 & $\sim$13$^{\prime\prime}$ & 3.0$\pm$1.5 & 0.2 & --  & [3] \\
J1834.9--0846 & 2.48 &  14 & 2.1$\times$10$^{34}$ & 5 & 70--150$^{\prime\prime}$  & 3.5$\pm$0.6 & 0.7  & W41? & [4] \\
 1E~1547.0--5408 & 2.07 & 32 & 1.0$\times$10$^{35}$ & 0.7 & 45$^{\prime\prime}$ & 3.4$\pm$0.4 & 0.01 &   G327.2--0.1 & [5] \\
 \hline
\end{tabular}
{\it Notes:} 
For J1846--0258, the small-scale PWN features (Fig.~2) show variability associated with the 2006 outburst.
For Swift J1834.9--0846, the photon index corresponds to pre-outburst, with a post-outburst value of 
$\Gamma$=3.2$^{+0.7}_{-0.6}$ (depending on the background model). 1E~1547.0--5408 is included for completeness (see text for details).
[1] Gonzalez \& Safi-Harb 2003, Safi-Harb \& Kumar 2008; [2]  Kumar \& Safi-Harb 2008, Ng et al. 2008;
[3] Rea et al. 2009, see also these proceedings; [4] Younes et al. 2012; [5] Vink \& Bamba 2009, Olausen et al. 2011.\\
\end{table}

\section{Conclusions and future prospects}
The pulsar/SNR community has gone a long way since the discovery of the neutron and the prediction of neutron star-SNR association some 80 years ago, followed by the discovery of pulsars 44 years ago. 
The nebulae blown by their relativistic winds offer a unique astrophysical laboratory for pulsar discovery
and for probing their outflows, environment, and diversity. 
The recent discoveries of PWNe around high-B pulsars are opening a new window to address their link to the more classical RPPs.

Many questions remain to be answered. In particular we don't know whether (all) high-B neutron stars
should power steady PWNe, and if so what powers their X-ray emission and what specifically distinguishes them from (or links them to) the other neutron star classes. 
Monitoring these sources, together with targeted deep X-ray and $\gamma$-ray observations, will help settle the puzzle about the driving mechanism and energy budget for their high-energy emission.
As well, the question on the (still) missing pulsars in many SNRs can be addressed with coordinated radio and high-energy studies and pulsation searches, especially in the X-ray detected PWNe.

We hope for many more years for the currently operating X-ray missions, in particular $Chandra$ as it has been instrumental in detecting 
PWNe and localizing previously missed pulsars.
The hard X-ray (5--80 keV) capabilities of $NuSTAR$, successfully launched in June 2012, the broadband coverage (0.3-600 keV) of \textit{ASTRO-H} slated for launch in 2014,
and the MeV to TeV coverage with existing and upcoming gamma-ray facilities, will provide a new window to the 
high-energy studies of these fascinating objects.

\acknowledgments\vspace{-0.5ex}
I greatly acknowledge the support of 
the CRC program, 
NSERC, CITA, CFI, and CSA.
This review wouldn't be possible without the contributions by PSR and SNR researchers, many not listed
here due to limited space. I am indebted to Zaven Arzoumanian for help with Fig.~1 which made use of data from the ATNF pulsar catalog and the McGill magnetars catalog.
Special thanks to the organizers for the invitation to a stimulating symposium and to  Joeri van Leeuwen for his cooperation in the editing process,
and best wishes to Dick  Manchester for a milestone birthday$!$


\begin{thebibliography}{}

\bibitem[]{arz11}
Arzoumanian, Z. et al.
 \textit{ApJ}, 739, 39.

\bibitem[]{BZ34}
Baade, W. \& Zwicky, F. 1934, \textit{Proc. Nat. Acad. Sci.}, 20, 254.
\bibitem[]{}
Camilo, F. 2008, \textit{Nature Physics}, 4, 353.

	
\bibitem[]{chadwich32}
Chadwick, J. 1932,  \textit{Proceedings of the Royal Society of London, Series A},
136, 692.

\bibitem[FerrandSSH12]{ferrandssh12}
Ferrand, G. \& Safi-Harb, S. 2012, \textit{Adv. Sp. Res.}, 49, 1313 (arXiv:1202.0245).

\bibitem[]{gaenslerslane2006}
Gaensler, B. M. \& Slane, P. O. 2006, \textit{ARAA}, 44, 17.


\bibitem[]{gaensler95}
Gaensler, B. M. et al. 2005, \textit{Nature}, 434, 1104.

\bibitem[]{gavriil08}
Gavriil, F. et al. 2008, \textit{Science}, 319, 1802.

\bibitem[]{gold68}
Gold, T. 1968, \textit{Nature}, 218, 731.

\bibitem[]{gonzssh03}
Gonzalez, M. E.,  \& Safi-Harb, S. 2003, \textit{ApJ}, 591, L143.


\bibitem[]{evghalpern08}
Gotthelf, E. V. \& Halpern, J. P. 2008, in \textit{AIP Conference Proceedings}, Volume 983, 320.

\bibitem[]{harding99}
Harding, A., Contopoulos, I., Kazanas, D. 1999, \textit{ApJ}, 525, L125

\bibitem[]{hewish68}
Hewish, A., Bell, S. J., Pilkington, J. D. H., Scott, P. F., Collins, R. A. 1968, \textit{Nature}, 217, 709.
	
\bibitem[]{KP2010}
Kargaltsev, O. \& Pavlov, G. G.  2008, \textit{AIP Conference Proceedings}, Volume 983, 171.


\bibitem[]{keane11}
Keane, E. F. et al. 2011, \textit{MNRAS}, 415, 3065.
	
\bibitem[]{KSSH08}
Kumar, H. S. \& Safi-Harb, S. 2008, \textit{ApJ}, 678, L43.

\bibitem[]{kumar12}
Kumar, H. S., Safi-Harb, S. \& Gonzalez, M. E. 2012, \textit{ApJ}, 754, 96.

\bibitem[]{large68}
Large, M. I., Vaughan, A. E., Mills, B. Y. 1968, \textit{Nature}, 220,  340.

\bibitem[]{lovelace68}
Lovelace, R. B., Sutton, J. M., Craft, H. D.  1968, \textit{IAU Circ.}, 2113, 1.

\bibitem[]{mereghetti08}
Mereghetti, S. 2008,  \textit{A\&ARv}, 15, 225.

\bibitem[]{ngkaspi11}
Ng, C.-Y. \& Kaspi, V. M. 2011, \textit{AIP Conference Proceedings}, Volume 1379, 60.
	
\bibitem[]{ng08}
Ng, C.-Y., Slane, P. O., Gaensler, B. M. \&  Hughes, J. P. 2008, \textit{ApJ}, 686, 508.

\bibitem[]{olausen11}
Olausen, S. A.  et al. 2011, \textit{ApJ}, 742, 4.

\bibitem[]{pacini67}
Pacini, F. 1967, \textit{Nature},  216, 567.

\bibitem[]{pacsal73}
Pacini, F. \& Salvati, M. 1973, \textit{ApJ}, 186, 249.

\bibitem[]{rea09}
Rea, N. et al. 2009, \textit{ApJ}, 703, L41.


\bibitem[]{reesgunn74}
Rees, M. J. \& Gunn, J. E. 1974, \textit{MNRAS}, 167, 1.

\bibitem[Safi-Harb (2012)]{SH12}
{Safi-Harb, S.} 2012,
\textit{AIP Conference Proceedings}, Gamma2012 symposium
 (arXiv:1210.5406).

\bibitem[]{sshkumar08}
Safi-Harb, S. \& Kumar, H. S. 2008, \textit{ApJ}, 684, 532.

\bibitem[]{staelin68}
Staelin, D.H. \&  Reifenstein, E.C. 1968, \textit{Science}, 162, 1481-1483.
 

\bibitem[]{vinkbamba09}
Vink, J. \& Bamba, A. 2009, \textit{ApJ}, 707, L148.

\bibitem[]{W11}
Weltevrede, P., Johnston, S. \& Espinoza, C. M. 2011, \textit{MNRAS},
411, 1917.

 \bibitem[]{younes12}
Younes, G. et al. 2012,
\textit{ApJ}, 757, 39.
 
\end{thebibliography}
\end{document}